\UseRawInputEncoding
\documentclass[aps,prb,twocolumn,superscriptaddress,showpacs,preprintnumbers]{revtex4-1}
\usepackage[colorlinks,bookmarks=false,citecolor=blue,linkcolor=red,urlcolor=blue]{hyperref}
\usepackage{physics}
\input pdfcolor.tex
\usepackage{colortbl,amsthm,txfonts}
\usepackage{verbatim}
\usepackage{graphicx}
\usepackage{epsfig}
\usepackage{dcolumn}
\usepackage{bm}

\usepackage{amsmath,amssymb}

\makeatletter
\renewcommand*\env@matrix[1][*\c@MaxMatrixCols c]{%
  
  \hskip -\arraycolsep
  \let\@ifnextchar\new@ifnextchar
  \array{#1}}
\makeatother

\usepackage{appendix}

\begin{document}

\title{Transport and spectroscopic signatures of a disorder stabilized metal in two-dimensional frustrated 
Mott insulators}
\author{Madhuparna Karmakar}
\email{madhuparna.k@gmail.com}
\affiliation{Department of Physics, Indian Institute of technology, Madras, Chennai-600036, 
India.}
\author{Nyayabanta Swain}
\email{nyayabanta@nus.edu.sg}
\affiliation{Centre for Quantum Technologies, National University of Singapore, Singapore 117543.}

\begin{abstract}
Frustrated Mott insulators such as, transition metal dichalcogenides present 
an ideal platform for the experimental realization of externally tuned insulator-metal transition. 
In this paper we present the first non perturbative numerical investigation of the disorder induced 
insulator-metal transition in a two-dimensional frustrated Mott insulator. Our approach is 
generic and captures the essential physics of Mott insulator-metal transition in geometrically frustrated 
lattices. For concreteness, we have compared our results with the experimental observations on 
copper (Cu) intercalated 1T-TaS$_{2}$. Based on the magnetic, spectroscopic and transport 
signatures we have mapped out the thermal phase diagram of Cu intercalated 1T-TaS$_{2}$ and established 
that over a regime of moderate disorder strength this material hosts an antiferromagnetic metal. Moreover, 
the insulator-metal transition in this system is not tied to the loss of magnetic correlations, thereby 
giving rise to two quantum critical points.  The emergent non Fermi liquid metal is governed by resilient 
quasiparticles, that survive as the relevant low energy excitations even after the break down of the Fermi 
liquid description. The transport and spectroscopic signatures discussed in this letter are expected to 
serve as important benchmarks for future experiments on this and related class of materials.
\end{abstract}

\date{\today}
\maketitle
\section{Introduction}
Interplay of interaction and disorder in systems with strong electron-electron correlation has always been a 
subject of fundamental interest in condensed matter physics \cite{abrahams_rmp2001,kravchenko_rpp,valles_book}. 
In the non interacting limit, an infinitesimal disorder in known to lead to Anderson localization in one and 
two dimensions,  while a three dimensional system undergoes a metal-insulator transition (MIT) at a 
critical disorder strength \cite{anderson1958,abrahams1979}. 
In the opposite limit are the Mott insulators wherein strong electron-electron interaction 
leads to localization at commensurate electron filling \cite{mott_book}. 
At low temperatures, Mott insulators are often associated with magnetic ordering, 
low energy spin excitations and gapped charge excitations \cite{mott1949,fazekas_lecturenotes}. 
The outcome of the interplay between these two localizing tendencies was suggested to 
stabilize a metallic phase between the Mott and the Anderson 
insulating phases, leading to an insulator-metal-insulator transition, in two-dimensions (2D)
\cite{finkelstein1983,finkelstein2_1983}. 
The first possibility of a Mott insulator-metal transition (IMT) in 2D was suggested based 
on the renormalization group (RG) theory by Finkel'stein \cite{finkelstein1983,finkelstein2_1983},  
while the first experimental signature of the same was observed in the transport data of high mobility silicon metal 
oxide semiconductor field effect transistor \cite{kravchenko1994,kravchenko_rpp}. 

\begin{figure}[b]
\begin{center}
\includegraphics[height=6.5cm,width=7.5cm,angle=0]{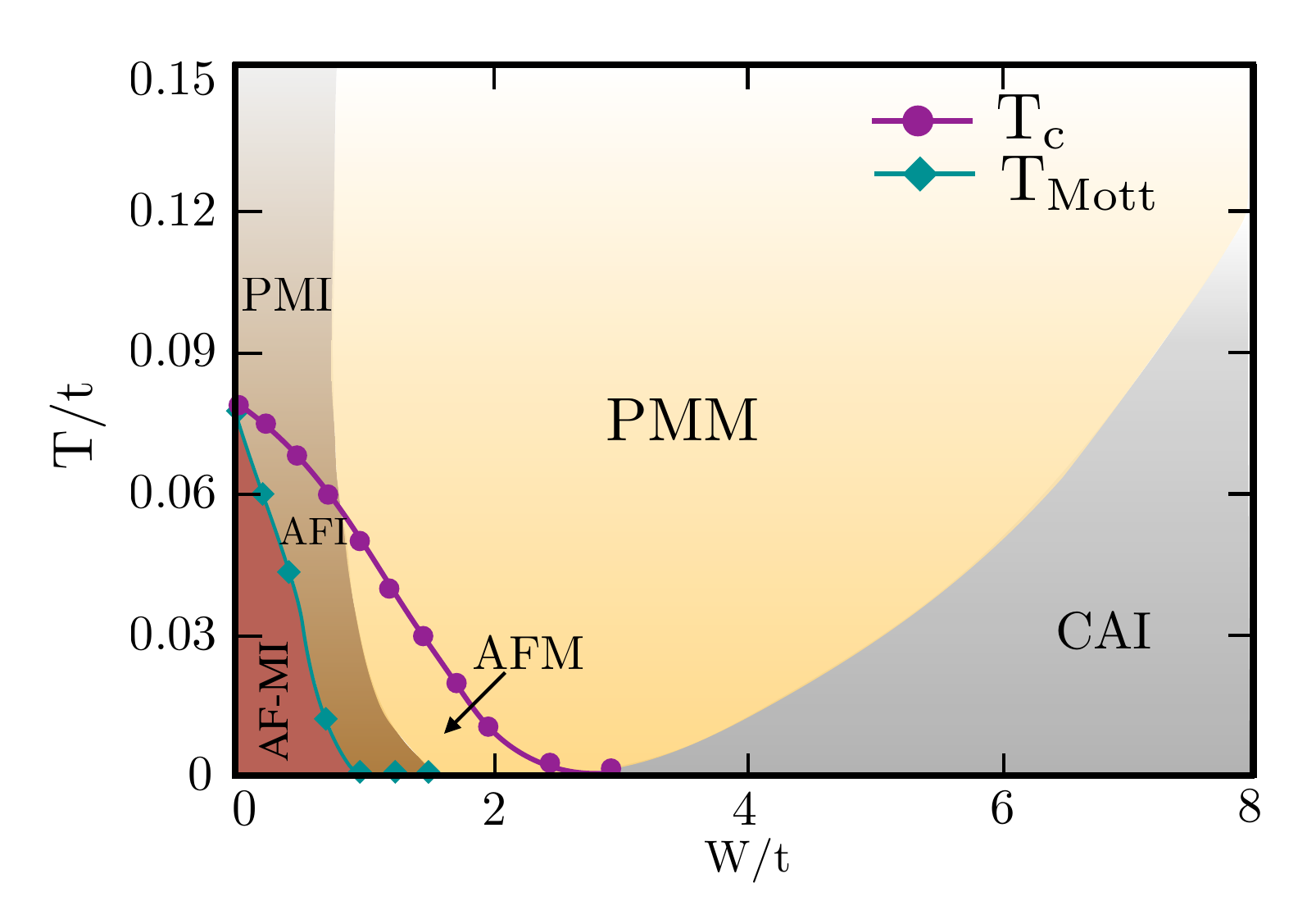}
\caption{\label{fig1}
Color online: Phase diagram of Anderson-Hubbard model on a triangular lattice in the 
disorder-temperature ($W-T$) plane at $U=6t$.
Different phases of this model are antiferromagnetic gapped Mott insulator (AF-MI), antiferromagnetic 
gapless insulator (AFI), antiferromagnetic metal (AFM), paramagnetic insulator (PMI), paramagnetic metal (PMM)
and the non magnetic correlated Anderson insulator (CAI). The thermal scales $T_{Mott}$ and $T_{c}$ indicate 
the temperatures where the Mott gap and the magnetic correlations vanish,  respectively. See text for 
detail characterization of the phases.}
\end{center}
\end{figure}
Disorder induced IMT on the bipartite square lattice is an well investigated subject, both at the ground state and at finite temperatures. 
The range of applied theoretical tools include,  inhomogeneous mean field theory \cite{heidarian2004,trivedi2020},
dynamical mean field theory (DMFT) and its variants \cite{byczuk2005,aguiar2008,oliviera2014,aguiar2009,semmler2010,tanaskovic2003,lee2016,song2008,kawakami2008,merino2009,miranda2020}, determinant quantum Monte Carlo (DQMC) \cite{srinivasan2003,otsuka2000,chiesa2008}, exact diagonalization \cite{kotlyar2001}, 
typical medium theory \cite{braganca2015} and its extensions \cite{ekuma2015}, auxiliary field quantum 
Monte Carlo with static path approximation \cite{dagotto2017} etc.
In the clean limit this system is a Mott insulator with a Neel antiferromagnetic order 
of the local moments. Disorder brings electron itinerancy back into the picture, 
thereby allowing the competition between insulating tendency of the localized moments and the metallicity of 
the itinerant electrons. The outcome of this competition is a metallic regime which lacks any magnetic ordering, 
at intermediate disorder. Notably, the system hosts a single quantum critical point (QCP)  corresponding 
to the IMT; the onset of metallicity is accompanied by the loss of (quasi) long range magnetic order 
\cite{dagotto2017}.

The primary difficulty in the experimental realization of IMT is to find suitable Mott insulators in 
which disorder can be introduced in a controlled manner without changing the carrier concentration. 
An avenue to circumvent this difficulty is the recently discovered 2D transition metal 
dichalcogenide (TMD) and related systems, which allows for the independent control over lattice filling 
and electron-electron correlations \cite{chowdhury2021,pashupathy2021}. 
One such candidate material is 1T-TaS$_{2}$, which is a geometrically 
frustrated system with a rich phase diagram \cite{fazekas1980,aryanpour2006,wilson1975,sipos2008,xu2010}.
 Recent experiments on intercalation of 1T-TaS$_{2}$ with potassium (K)
\cite{zhu2019}, copper (Cu) \cite{trivedi2014} and iron (Fe) \cite{niu2020}
have brought forth the possibility of IMT and a disorder stabilized ``magnetic'' metal in this material; 
 the origin of the same is currently debated \cite{lee2019,zitko2019,zhu2019}. 

Geometric frustration dictates the low temperature magnetic state both in clean and 
disordered systems \cite{balents_review2010,kanoda2020}. In presence of frustration 
the (quasi) long range magnetic order is not tied to the IMT and the magnetic correlations survive even after 
the Mott gap collapses. The system thus hosts different QCPs corresponding to the IMT and the loss of magnetic order. 
Multiple QCPs is a generic feature of geometrically frustrated lattices and has been observed 
in recent experiments on 2D TMDs \cite{pashupathy2021}. This is in sharp contrast with the bipartite lattices wherein 
IMT is tied to the loss of the magnetic order, leading to a single QCP \cite{dagotto2017}. Note 
that at low temperatures the magnetic order is closely tied to the lattice geometry as long as the disorder strength 
is comparable to the electron correlation.

In this paper we report the first non perturbative theoretical study of disorder induced IMT in 2D 
frustrated Mott insulators. While our approach and results are generic for such systems, for concreteness 
we present the example of Cu intercalated 1T-TaS$_{2}$ and compare our results with the relevant experimental 
observations made on this material. The tool of our choice is auxiliary field quantum Monte Carlo technique with 
static path approximation (SPA) \cite{evenson1970}, which has been extensively 
utilized in the context of magnetic and superconducting systems \cite{meir2007,tarat_disorder,rajarshi,nyayabanta_pyrochlore,nyayabanta_chk,mpk_imb,mpk_mass,lieb_strain} (see appendix). 
Recently SPA has also been used to investigate disorder induced IMT in square lattice; the results obtained 
therein are in agreement with the existing understanding of IMT in unfrustrated lattices \cite{heidarian2004,trivedi2020,byczuk2005,aguiar2008,oliviera2014,aguiar2009,semmler2010,tanaskovic2003,lee2016,song2008,kawakami2008,merino2009,miranda2020,kotlyar2001,braganca2015,ekuma2015,denteneer1999,chiesa2008}.
 
Our key results on disorder induced IMT in 2D frustrated Mott insulators are as follows:       
$(i)$~We map out the generic thermal phase diagram to show the disorder induced IMT in 2D frustrated 
Mott insulator, wherein a metallic phase is stabilized over a regime of intermediate disorder potential. 
$(ii)$~Owing to the geometric frustration the system hosts separate QCPs corresponding to the IMT and loss 
of (quasi) long range magnetic order. 
$(iii)$~The disorder stabilized emergent {\it antiferromagnetic} metal is a non Fermi liquid (NFL), 
characterised by resilient quasiparticles (QP), as attested via the transport and spectroscopic signatures. 
$(iv)$~The metallic phase exhibits a thermal crossover between 
NFL and bad metallic phases as a re-entrant thermal transition, characterized by a change in the sign of 
$d\sigma_{dc}/dT$ (T is temperature and $\sigma_{dc}$ is dc conductivity, discussed later). 
$(v)$~Our results provide the first microscopic description of the existing experimental observations on Cu 
intercalated 1T-TaS$_{2}$ and serves as benchmarks for related class of materials. 

\section{Model and method}
We model the disordered 2D frustrated Mott insulator as the Anderson Hubbard model on a triangular 
lattice \cite{fazekas_lecturenotes}, which reads as,

\begin{eqnarray}
H & = & \sum_{\langle ij\rangle, \sigma}t_{ij}(c_{i\sigma}^{\dagger}c_{j\sigma} + h . c)
+ \sum_{i,\sigma}(W_{i}-\mu)\hat n_{i\sigma} + U\sum_{i}\hat n_{i\uparrow}\hat n_{i\downarrow} \nonumber 
\end{eqnarray}

\noindent here, $t_{ij}=-t$,  for the nearest neighbor hoppings on an isotropic triangular lattice 
(coordination number $z=6$). 
$t=1$ is set as the reference energy scale. $U > 0$ is the on-site repulsive Hubbard interaction. 
The site dependent disorder is introduced in the system via $W_{i}$ which is randomly selected from 
an uniform distribution  $[+W/2, -W/2]$. We work at half filling and the chemical potential $\mu$ is 
adjusted to achieve the same. 

We make this model numerically tractable by decomposing the interaction 
term via Hubbard Stratonovich (HS) decomposition \cite{hubbard1959,schultz1990}, which introduces a vector 
${\bf m}_{i}(\tau)$ and a scalar $\phi_{i}(\tau)$ (bosonic) auxiliary field 
at each site, where the former couples to the spin and the later to the charge. 
We next drop the time dependence of these auxiliary fields and treat them as ``classical'' fields ${\bf m}_{i}$ and 
$\phi_{i}$. The thermal fluctuations in ${\bf m}_{i}$ are retained completely but $\phi_{i}$ is treated at the 
saddle point level, $\phi_{i}\rightarrow \langle \phi_{i}\rangle = \langle n_{i}\rangle U/2$. These 
approximations lead to a coupled spin-fermion model wherein the fermions move in the 
spatially fluctuating background of ${\bf m}_{i}$. 
The equilibrium configurations of $\{{\bf m}_{i}\}$ are generated via Monte Carlo (MC) simulation 
and the different correlators are computed on these equilibrium configurations.
Technical details of the method are discussed in the appendix.  
\begin{figure}
\begin{center}
\includegraphics[height=8.8cm,width=8.7cm,angle=0]{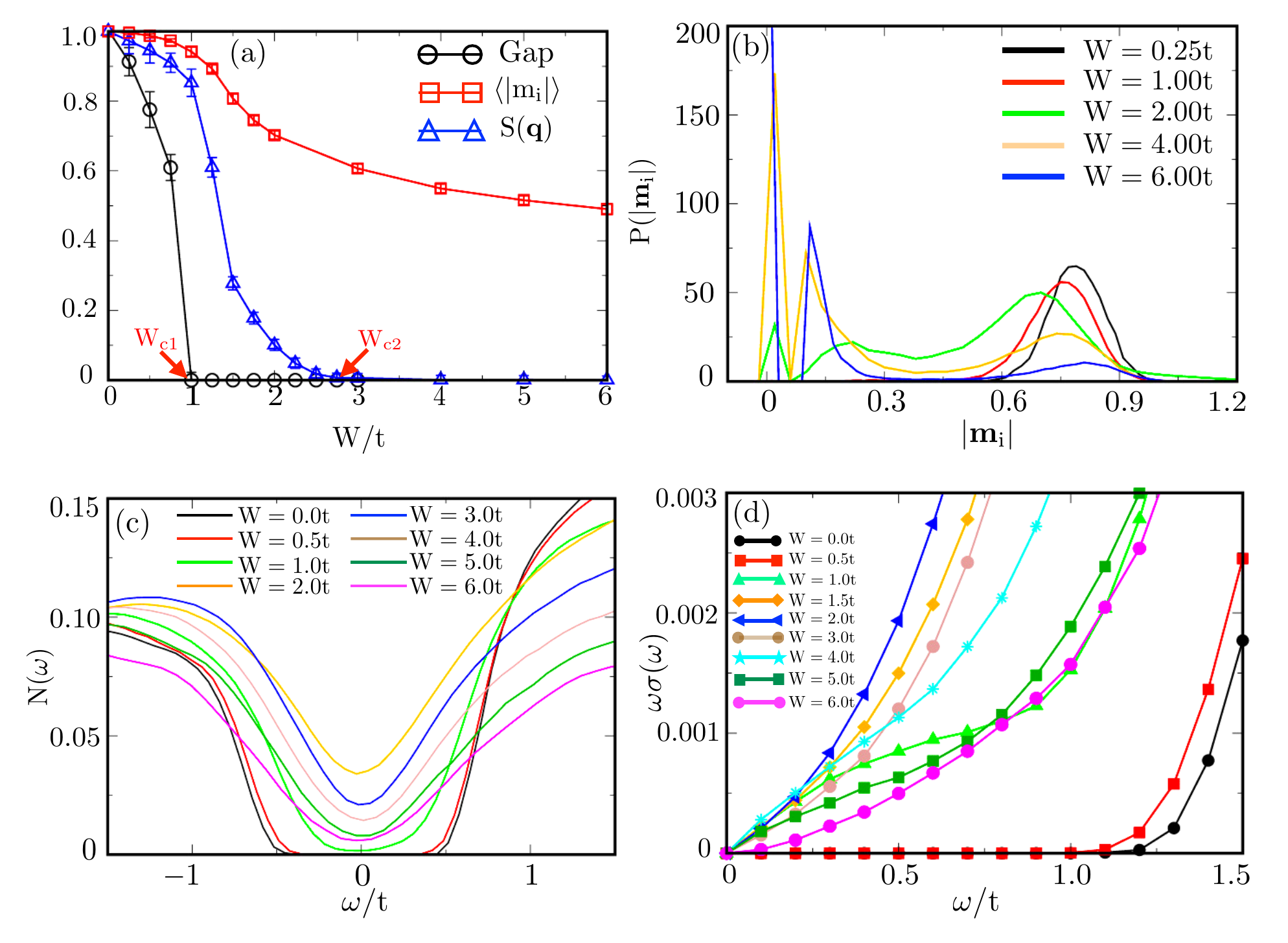}
\caption{Color online: (a) Disorder dependence of normalized single particle energy gap, (Gap), 
normalized magnetic moment amplitude ($\langle |m_{i}| \rangle$) and normalized magnetic structure factor peak 
($S({\bf q})$), at $T=0$.  The error bars correspond to the cumulative error arising out of MC and disorder 
averaging of the observables. (b) Distribution of magnetic moments (P($\vert {\bf m}_{i}\vert$)) at $T=0$ for 
different disorder potentials. Note the transfer of weight towards $\vert {\bf m}_{i}\vert = 0$ with 
increasing disorder, indicating the spatial fragmentation of the system into puddles with suppressed 
amplitude of local moments, followed by the percolation of these puddles. (c) Single particle DOS 
($N(\omega)$) as function of disorder at $T=0$. (d) Scaled optical conductivity ($\omega\sigma(\omega)$) 
(in appropriate units) plotted over a small range of low frequencies, for different disorder potential, at $T=0$.}
\label{fig2}
\end{center}
\end{figure}

\begin{figure*}
\begin{center}
\includegraphics[height=4.5cm,width=16.0cm,angle=0]{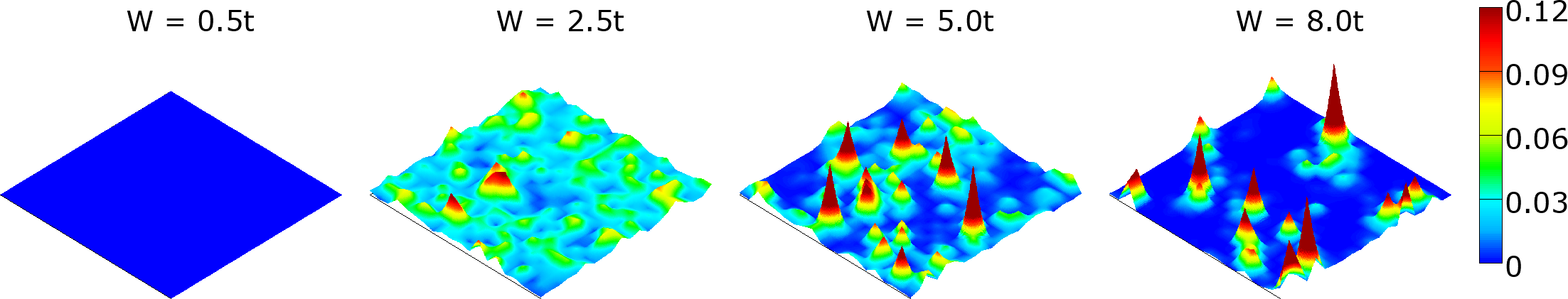}
\caption{Color online: Spatial maps of local density of states ($N_{i}(0)$) at the Fermi level at different disorder 
potential across the insulator-metal-insulator transition. While both and AF-MI and CAI are characterized by localized 
energy states, the AFM regime shows significant delocalization of the energy states (see text).}
\label{fig3}
\end{center}
\end{figure*}
\section{Results}
The different phases of this system are characterized by
the following indicators: 
\begin{itemize}
\item{Magnetic structure factor,
\begin{eqnarray}
S({\bf q}) & = & \frac{1}{N^{2}}\sum_{ij}\langle {\bf m}_{i}.{\bf m}_{j}\rangle e^{i{\bf q}.({\bf r}_{i}-{\bf r}_{j})}
\end{eqnarray}
where, ${\bf q}$ corresponds to the magnetic ordering wave vector and $N$ is the number of lattice sites. 
$\langle ...\rangle$ corresponds to MC configurational average.
}

\item{Single particle density of states (DOS),
\begin{eqnarray}
N(\omega) & = & (1/N)\sum_{\alpha}\langle \delta(\omega - \epsilon_{\alpha})\rangle
\end{eqnarray}
where, $\epsilon_{\alpha}$ are eigenvalues in a single equilibrium configuration.
}

\item{Single particle local density of states (LDOS),
\begin{eqnarray}
\rho_{i}(\omega) & = & \sum_{\alpha,\sigma}\langle \vert u_{\alpha,\sigma}^{i}\vert^{2} 
\delta(\omega - \epsilon_{\alpha})\rangle
\end{eqnarray}
where, $u_{\alpha,\sigma}^{i}$ is the eigenvector corresponding to the eigenvalue $\epsilon_{\alpha}$.
}

\item{Inverse participation ratio (IPR),
\begin{eqnarray}
IPR & = & \frac{\sum_{i, \alpha, \sigma}\vert u_{\alpha,\sigma}^{i}\vert^{4}}{[\sum_{i, \alpha, \sigma}\vert u_{\alpha, \sigma}^{i}\vert^{2}]^{2}}
\end{eqnarray}}
\item{Optical conductivity, calculated using the Kubo formula, 
\begin{eqnarray}
\sigma(\omega) & = & \frac{\sigma_{0}}{N}\sum_{\alpha, \beta}\frac{n_{\alpha}-n_{\beta}}{\epsilon_{\beta}-
\epsilon_{\alpha}}\vert \langle \alpha\vert J_{x}\vert \beta\rangle\vert^{2}\delta(\omega-(\epsilon_{\beta}-
\epsilon_{\alpha}))
\end{eqnarray}
}
where, the current operator $J_{x}$ is defined as, 
\begin{eqnarray}
J_{x} & = & -i\sum_{i,\sigma,\vec{\delta}}[\vec{\delta}t_{\vec{\delta}}c_{{\bf r}_{i},\sigma}^{\dagger}
c_{{\bf r}_{i}+\vec{\delta},\sigma}-h.c]
\end{eqnarray}
The dc conductivity ($\sigma_{dc}$) is the $\omega \rightarrow 0$ limit of $\sigma(\omega)$, 
$\sigma_{0}=\frac{\pi e^{2}}{\hbar}$ in 2D. $n_{\alpha} = f(\epsilon_{\alpha})$ is the Fermi 
function, and $\epsilon_{\alpha}$ and $\vert \alpha\rangle$ are respectively the single particle 
eigenvalues and eigenvectors of $H_{eff}$ in a given background of $\{{\bf m}_{i}\}$.

\item{Spectral line shape,
\begin{eqnarray}
A({\bf k}, \omega) & =& -(1/\pi){\mathrm{Im}} G({\bf k}, \omega)
\end{eqnarray}
where, $G({\bf k}, \omega) = lim_{\delta \rightarrow 0} G({\bf k}, i\omega_{n})\vert_{i\omega_{n}\rightarrow \omega + i\delta}$. 
$G({\bf k}, i\omega_{n})$ is the imaginary frequency transform of $\langle c_{\bf k}(\tau)c_{\bf k}^{\dagger}(0)\rangle$.
}
\end{itemize}
The results presented here corresponds to a system size of $24 \times 24$
and are found to be robust against finite system size effects (see appendix). 
The on-site Hubbard repulsion is selected to be $U=6t$ which is close to the Mott boundary 
of the triangular lattice in the clean limit \cite{imada2002,inoue2008,kawakami2009,becca2009,limelette2003,aryanpour2006,phillips2009,kotliar2004,kawakami2008,merino2009,sato2012,senechal2008,schmidt2010,ohta2017,becca2020,rajarshi,georges2021,dressel2021}. 
This choice of $U$ ensures that our parameter regime is well suited to capture the experimental observations 
on Cu intercalated 1T-TaS$_{2}$ \cite{trivedi2014,sipos2008}. The observables are averaged over 30 disorder 
realizations. 

\subsection{Phase diagram and thermal scales} 

In the clean limit, the half filled triangular lattice at $U=6t$ hosts a 120$^{\circ}$ spiral magnetic order with 
the ordering wave vector ${\bf Q}=\{2\pi/3, 2\pi/3\}$, in the ground state
\cite{imada2002,inoue2008,kawakami2009,becca2009,limelette2003,aryanpour2006,phillips2009,kotliar2004,kawakami2008,merino2009,sato2012,senechal2008,schmidt2010,ohta2017,becca2020,rajarshi}. 
The corresponding antiferromagnetic Mott insulator (AF-MI) phase undergoes thermal transition to a paramagnetic 
insulator (PMI) comprising of randomly oriented magnetic moments \cite{rajarshi}. 
Introduction of quenched random disorder alters this picture significantly and leads to a 
disorder induced IMT, as shown in Figure \ref{fig1}. 
We characterize the resulting phases based on the indicators shown in Figure \ref{fig2}. 
The regime of weak disorder ($0 < W \lesssim t$) is akin to the clean limit 
and hosts an AF-MI in the ground state, comprising of a robust spectral gap at the Fermi level. 
Increasing disorder leads to 
progressive accumulation of spectral weight at the Fermi level such that the Mott gap collapses for $W \ge t$,
even though the magnetic order survives. The collapse of the Mott gap sets the first QCP 
($W_{c1} \sim t$) of the system, while a second QCP ($W_{c2} \sim 2.75t$) 
is set by the loss of the magnetic order. 
We show these QCPs in Figure \ref{fig2}(a), as disorder dependence 
of Mott gap and magnetic structure factor peaks, respectively. In addition, we show 
the disorder dependence of the average local moment amplitude ($\langle |m_{i}| \rangle$), which 
survives even at strong disorder. The system does not undergo any spontaneous symmetry 
breaking across the disorder induced quantum phase transitions. The order of these 
transitions are reminiscent of weak first order transition observed in the clean 
limit \cite{moeller1999,park2008,casey2003}. 

The distribution of the average magnetic moment amplitude (P($|{\bf m}_{i}|$)) 
is shown in Figure \ref{fig2}(b). At weak disorder P($|{\bf m}_{i}|$) is unimodal indicating 
a homogeneous distribution of local moments with a mean amplitude of $\bar  m$. Increase in disorder results 
in progressive fragmentation of the magnetic state indicated by the broadening of the peak. Along 
with $\bar m$ there is accumulation of weight at $|{\bf m}_{i}|=0$ which indicates that 
the system comprises of regions with vanishing local moment amplitude. 
The intermediate disorder regime shows that the contributions from 
$\bar  m$ and $|{\bf m}_{i}|=0$ are comparable. On further increasing disorder the regimes with suppressed 
magnetic moments percolate through the system,  thereby destroying the magnetic order. The corresponding 
P($|{\bf m}_{i}|$) shows that the weight is now largely accumulated at $|{\bf m}_{i}|=0$.

\begin{figure*}
\begin{center}
\includegraphics[height=6.8cm,width=14.5cm,angle=0]{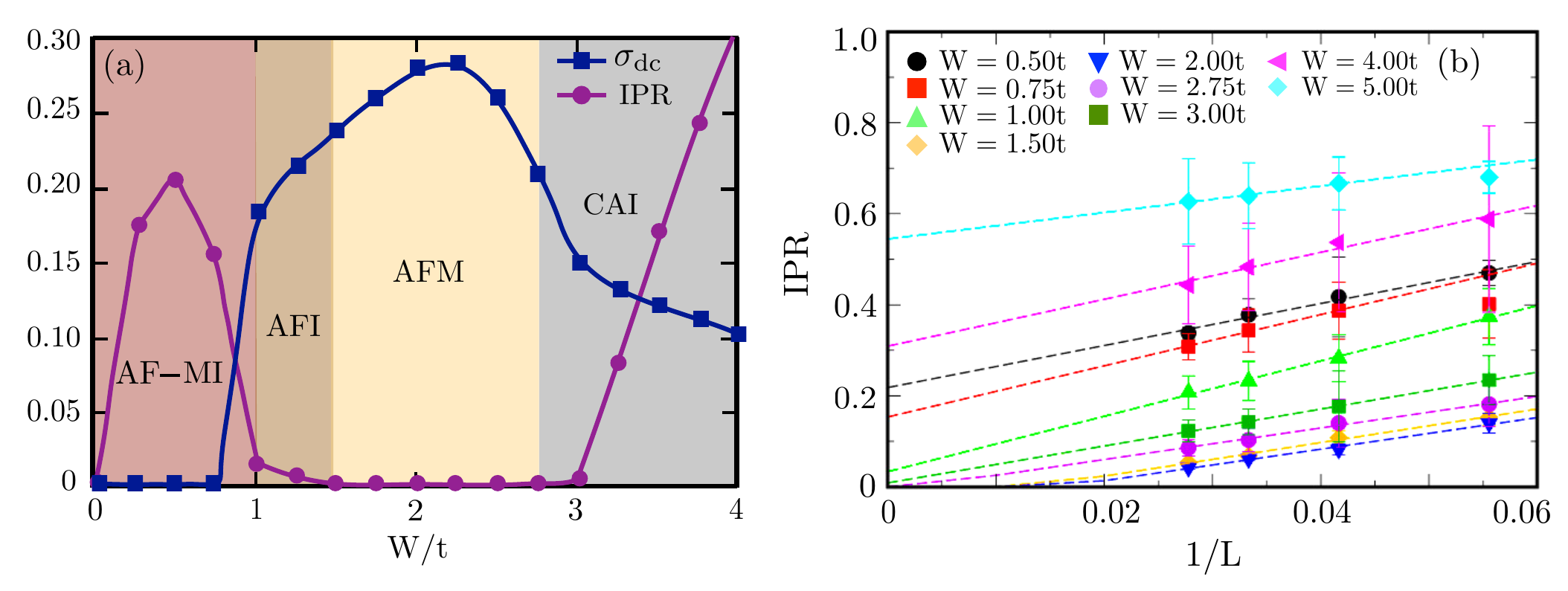}
\caption{Color online:  (a) Disorder dependence of Inverse participation ratio (IPR) at the Fermi level in the thermodynamic limit and dc conductivity ($\sigma_{dc}$) 
at $L = 24$, at $T = 0$. (b) Finite system size scaling of IPR at the Fermi level at indicated disorder potentials. The points represent the numerical data. Dashed lines 
are linear fit to the data. The IPR vanishes for metal and extrapolates to a finite 
value for the insulating phase as $L \rightarrow \infty$.}
\label{fig4}
\end{center}
\end{figure*}

In Figure \ref{fig2}(c) we show the single particle DOS at the ground state for a range of disorder potentials. 
While the AF-MI comprises of a robust spectral gap, the antiferromagnetic insulator (AFI), 
antiferromagnetic metal (AFM) and correlated Anderson insulator (CAI) shown in Figure  \ref{fig1} are gapless 
phases and can not be distinguished based on the single particle DOS alone. The distinction 
is made based on the transport properties of the system.
Transport signature of the system is shown in Figure \ref{fig2}(d) in terms of the scaled optical conductivity 
($\omega\sigma(\omega)$). For $W \le W_{c1}$, $\omega\sigma(\omega)=0$, corresponding to a gapped
 AF-MI. Over a narrow regime $W_{c1} \le W < 1.50t$, $\omega\sigma(\omega) \rightarrow 0$ as 
$\omega \rightarrow 0$, {\it nonlinearly}, indicating an insulating phase \cite{imry_book}. 
In the regime $1.50t < W \le W_{c2}$, $\omega \sigma(\omega) \rightarrow 0$ {\it linearly}, thus,  
$\sigma(\omega)$ is a constant at small $\omega$, corresponding to a metal. 

\subsection{Disorder induced localization}

The regime $W>W_{c2}$ of Figure \ref{fig1} corresponds to the CAI, characterized by
the absence of magnetic order, presence of gapless excitations and localized low energy
single particle eigenstates. In Figure \ref{fig3} we show the localization of 
the single particle states at the representative disorder strengths, in terms of the LDOS 
at the Fermi level. In the AF-MI phase 
LDOS is gapped, at $W=2.5t$ the system is an AFM, characterized by the delocalized energy states as shown in 
the LDOS maps. The system is in the CAI phase both at $W=5t$ and at $W=8t$, and the 
corresponding LDOS maps show progressively robust localization of the energy states. At still stronger 
disorder (not shown here) the energy states at the Fermi level gets completely localized via Anderson 
localization phenomenon.

In Figure \ref{fig4}(a) we show the inverse participation ratio (IPR) and the dc conductivity 
($\sigma_{dc}$) at the ground state, that have been used to characterize the CAI phase.
The localization length ($\xi_{loc}$) of an eigenstate $\psi_{i, \alpha}$ is related to the IPR as, IPR $\propto \xi^{-2}_{loc}$; consequently,  increase 
in disorder is expected to reduce $\xi_{loc}$ and increase IPR. The disorder dependence 
of IPR at the Fermi level shown in Figure \ref{fig4}(a) corresponds to the results at the thermodynamic limit, 
determined based on the finite system size scaling analysis of IPR, shown in Figure \ref{fig4}(b). Our result 
suggests that for both AF-MI and CAI, the IPR increases with increasing disorder. The disorder regime 
$W_{c1} < W \le 1.50t$ corresponding to the AFI is weakly localized and has a small but finite IPR 
at the Fermi level. The intermediate disorder regime ($1.50t < W \le W_{c2}$) is anomalous and an increase in disorder 
leads to delocalization of the eigenstates. In the thermodynamic limit, as shown in Figure \ref{fig4}(a), 
IPR vanishes over this regime of intermediate disorder potential, as is expected from a metallic phase. 
The anomaly in IPR over the regime $W_{c1} \le W < W_{c2}$ is reflected in transport signatures as 
well, in terms of significant increase in $\sigma_{dc}$, attesting an AFI and a metallic phase. 

\subsection{Finite temperature scales}

\begin{figure}[b]
\begin{center}
\includegraphics[height=5.0cm,width=8.8cm,angle=0]{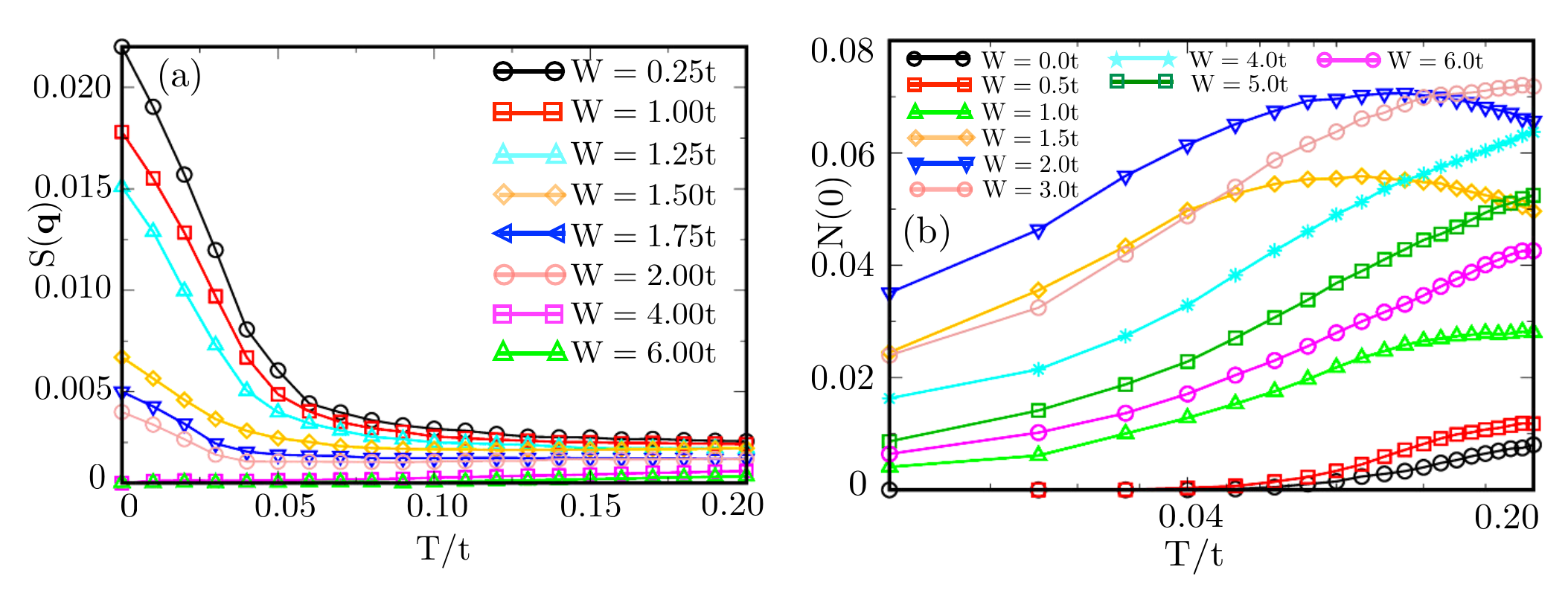}
\caption{Color online: (a) Thermal evolution of magnetic structure factor peak, $S({\bf q})$ and 
(b) single particle DOS at the Fermi level, $N(0)$, at different disorder potentials.}
\label{fig5}
\end{center}
\end{figure}

Figures \ref{fig5}(a)-(b) correspond to the indicators based on which the thermal scales shown in Figure  \ref{fig1}
are determined. Figure  \ref{fig5}(a) shows the thermal evolution of the magnetic structure factor peak ($S({\bf Q})$) 
at different disorder potentials. The point of inflection of these curves correspond to the $T_{c}$, 
which undergoes suppression with increasing disorder. Note that as per the Mermin-Wagner
theorem in 2D the system can not host any long range order at $T > 0$ \cite{mermin_wagner}.
The low temperature magnetic state observed in this system is (quasi) long range ordered; 
it comprises of magnetic correlations which decay as power law, unlike the exponential decay one would 
expect from a true disordered phase. Increase in temperature and the associated thermal fluctuations lead to 
progressive suppression of this ``algebraic long range order'' such that at high temperatures the correlations 
are short ranged and exponentially decaying. 

\begin{figure*}[t]
\begin{center}
\includegraphics[height=5.0cm,width=15.8cm,angle=0]{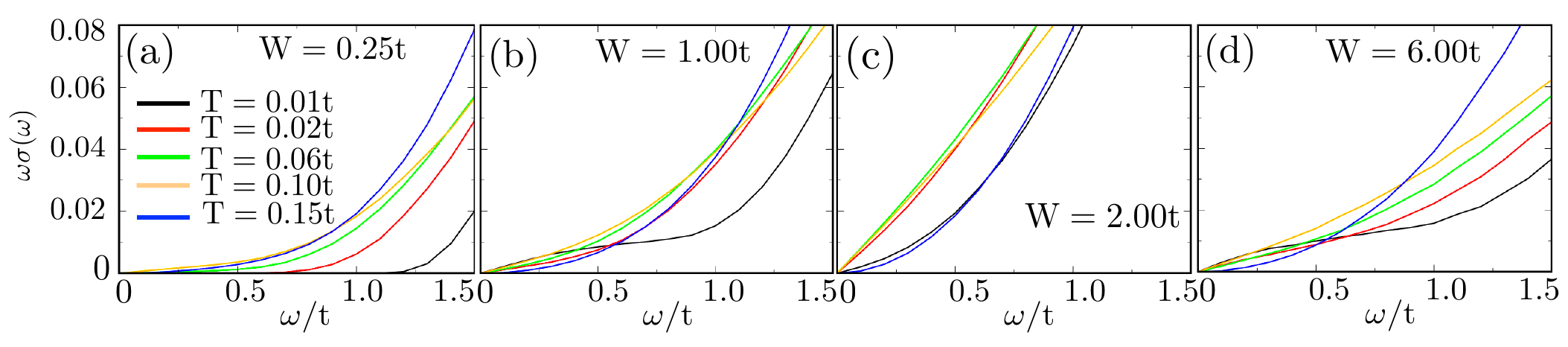}
\caption{Color online: Thermal evolution of scaled optical conductivity ($\omega \sigma(\omega)$)  at 
selected disorder cross sections. A nonlinear dependence of $\omega\sigma(\omega) \rightarrow 0$ as 
$\omega \rightarrow 0$ indicates an insulating behavior, while for a metal the dependence is linear.}
\label{fig6}
\end{center}
\end{figure*}

The thermal evolution of the single particle DOS at the Fermi level ($N(0)$) is shown in Figure \ref{fig5}(b), 
based on which $T_{Mott}$ is determined. For a given disorder 
potential, temperature leads to fluctuation of magnetic moments and accumulation of spectral weight  
at the Fermi level. The high temperature phase is thus a paramagnetic insulator (PMI) for $W < 1.50t$, and 
a paramagnetic metal (PMM) for $W \ge 1.50t$. Note that $dN(0)/dT$ undergoes a temperature dependent change 
in sign in the regime $W_{c1} \le W < W_{c2}$, indicating a weak ``re-entrant'' thermal transition.  

\begin{figure*}
\includegraphics[height=8.6cm,width=16.5cm,angle=0]{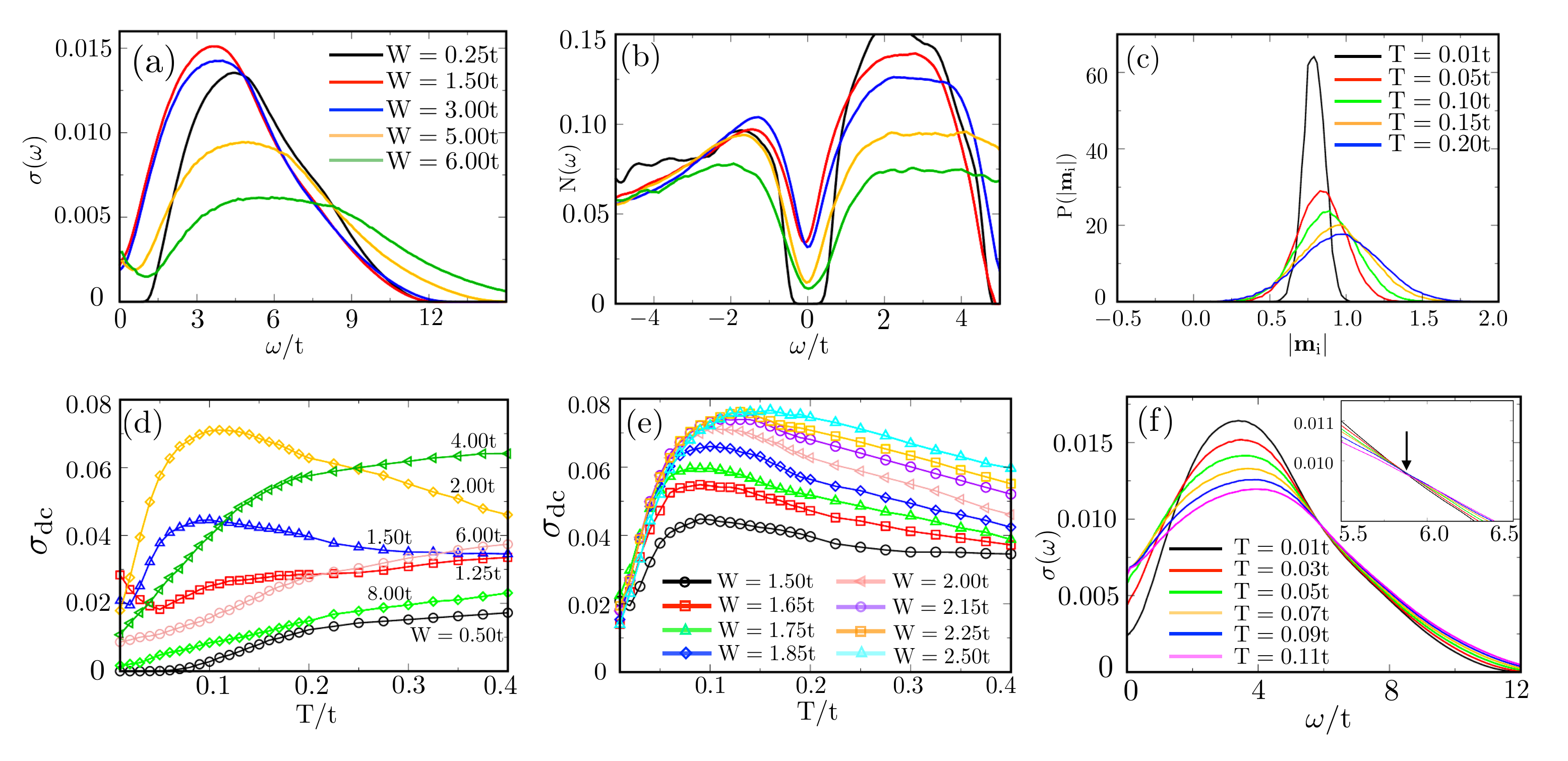}
\caption{\label{fig7}
Color online: (a) Optical conductivity ($\sigma(\omega)$) and (b) single particle DOS 
($N(\omega)$) at different disorder potential at $T=0.06t$.
(c) Thermal evolution of the magnetic moment distribution profiles (P$(|{\bf m}_{i}|)$) at $W=0.25t$.
(d) Thermal evolution of dc conductivity ($\sigma_{dc}$) at different disorder cross sections. The NFL 
regime at intermediate disorders is shown in detail in (e). (f) Thermal evolution of optical conductivity 
at $W=1.75t$. Inset: isosbestic crossing point ($\omega \sim 5.8t$) corresponding to thermal crossover 
from NFL to bad metallic phase.} 
\end{figure*}

The finite temperature metal-insulator boundaries of our phase diagram are determined based on the
behavior of $\omega \sigma(\omega)$ in the low frequency regime, shown in Figure \ref{fig6}.
At $W=0.25t$,  even at high temperatures the accumulation of 
weight in the low frequency regime is insignificant and the system continues to remain gapped. 
This high temperature regime corresponds to a PMI characterized by randomly oriented magnetic 
moments and a spectral gap at the Fermi level. At $W=t$ the system is in the AFI phase at low 
temperature which is signalled by the non linear behavior of
$\omega\sigma(\omega) \rightarrow 0$ as $\omega \rightarrow 0$. Increase in temperature leads 
to crossover of the system to a metallic phase and accordingly the low frequency behavior of 
$\omega\sigma(\omega)$ progressively becomes linear. The phase at $W=2t$ is an AFM, the corresponding 
$\omega\sigma(\omega) \rightarrow 0$ linearly as $\omega \rightarrow 0$. 
The metallic regime both at low and at high temperatures is characterized by the linear low frequency dependence of 
$\omega\sigma(\omega)$. The insulating CAI phase at $W=6t$ is once again signalled by the non linear behavior of 
$\omega\sigma(\omega) \rightarrow 0$ in the low frequency regime.
 For $T \gtrsim  0.1t$ the linear frequency dependence of $\omega \sigma(\omega)$ 
signals a finite temperature insulator-metal (CAI-PMM) crossover. 

\subsection{Non Fermi liquid metal and thermal crossover}

In order to obtain a deeper insight into the characteristics of the metallic regime 
at intermediate disorder potentials we next analyze the high frequency behavior of $\sigma(\omega)$, 
as shown in Figure \ref{fig7}(a), at $T=0.06t$.   Based on $\sigma(\omega)$ the different 
disorder regimes are characterized as follows:
$(i)$~ AF-MI at weak disorder is gapped at the Fermi level, $\sigma(\omega)$ contains displaced 
Drude peak (DDP) at $\omega \neq 0$; $(ii)$~ moderate disorder closes the gap via spectral weight 
accumulation and DDP shifts to low frequencies, indicating an IMT; $(iii)$~ strong disorder leads 
to CAI phase and accordingly DDP shifts back to higher $\omega$. 
The corresponding single particle DOS is shown in Figure \ref{fig7}(b). As the system 
undergoes transition from the AF-MI to the AFM state with increasing disorder, $N(\omega)$ 
accumulates large spectral weight at the Fermi level, indicating a {\it pseudogap} behavior. 
Note that similar pseudogap behavior have been observed in DQMC studies of disorder induced IMT on 
square lattice \cite{chiesa2008}.
The transition from the AFM to the CAI phase is signalled by the loss of spectral weight at the Fermi level 
as shown in Figure \ref{fig7}(b).

\begin{figure}[t]
\begin{center}
\includegraphics[height=4.5cm,width=8.5cm,angle=0]{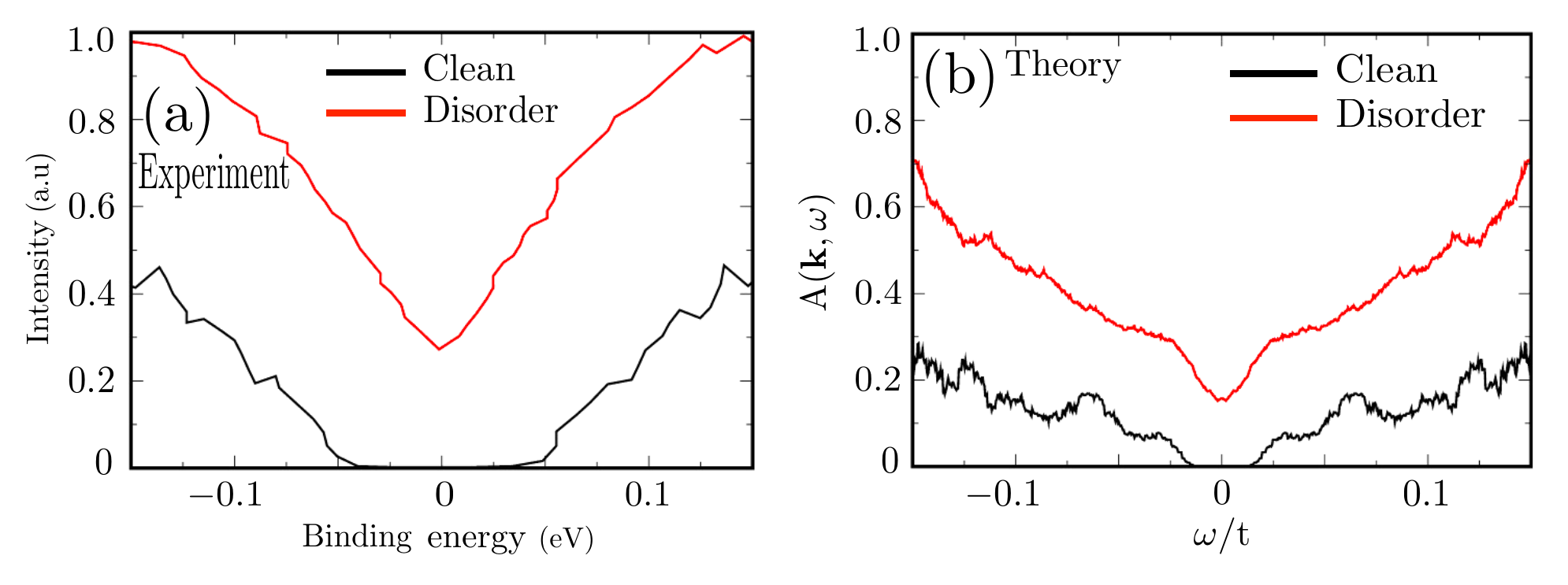}
\caption{Color online: (a) Symmetrized experimental energy distribution curve (EDC) determined 
from ARPES measurement on clean and Cu intercalated 1T-TaS$_{2}$, at the $\Gamma$-point; data extracted 
from \cite{trivedi2014}. (b) Symmetrized spectral line shape ($A({\bf k}, \omega)$) as determined numerically 
at a low temperature for the clean ($W = 0$) and disordered ($W = 1.75t$) cases at the $\Gamma$-point of the 
Brillouin zone.}
\label{fig8}
\end{center}
\end{figure}

In Figure \ref{fig7}(c) we present the temperature dependence of magnetic 
moment distribution (P($|{\bf m}_{i}|$)), at a selected disorder potential of $W=0.25t$. 
The generic effects of temperature at any disorder cross section are, 
$(i)$~ suppression of the peak height of P$(|{\bf m}_{i}|)$ and $(ii)$~ shifting of ${\bar m}$ towards 
larger values, with increasing temperature. This thermal dependence of $\vert{\bf m}_{i}\vert$ 
arises because temperature destroys angular correlations and therefore suppresses 
P$(|{\bf m}_{i}|)$; moreover, thermal fluctuations enhance ${\bar m}$ but lead to 
a broader distribution showing that the underlying state is not magnetically ordered. Had the 
underlying state been ordered, this increase in ${\bar m}$ 
would have strengthened the order and made the spectral gap robust. 
However, in the absence of any order,  magnitude fluctuations and loss of angular correlation 
leads to accumulation of spectral weight at the Fermi level. This increases $N(0)$ and gives rise to 
a pseudogap. At even higher temperatures the increase in ${\bar m}$ slows down as 
shown in Figure \ref{fig7}(c), also the orientational fluctuations almost saturates; this leads to loss of spectral 
weight at the Fermi level and $N(0)$ undergoes suppression.

From the low frequency ($\omega \rightarrow 0$) behavior of $\sigma(\omega)$ we determine $\sigma_{dc}$, and 
show its thermal evolution in Figure \ref{fig7}(d). We distinguish the metallic and insulating phases 
primarily based on the sign of $d\sigma_{dc}/dT$, such that, $d\sigma_{dc}/dT < 0$ corresponds to metal while 
for an insulator $d\sigma_{dc}/dT > 0$. At high temperatures, $d\sigma_{dc}/dT > 0$ for the AF-MI ($W=0.5t$), 
AFI ($W=1.25t$) and CAI ($W=4t, 6t, 8t$) phases and $\sigma_{dc}$ increases monotonically. In the intermediate 
disorder regime ($W=1.5t, 2t$) representative of the AFM, $d\sigma_{dc}/dT$ undergoes a change in sign across 
a crossover temperature at which $\sigma_{dc}$ is maximum, say, $T_{IRM}$.  
In Figure \ref{fig7}(e) we explore the metallic phase at the intermediate disorder regime, in detail. 
The non-monotonic temperature dependence of $\sigma_{dc}$ and the corresponding deviation 
of resistivity ($\rho=1/\sigma_{dc}$) from the $\rho \propto T^{2}$ behavior at low temperatures is the 
signature of the underlying NFL state.
The ``pseudogap'' behavior of the single particle DOS (Figure \ref{fig7}(b)) at the Fermi level 
suggests that even though the Fermi liquid description breaks down in the NFL state, the low 
energy excitations can still be described via the QPs. For $T \neq 0$ the finite spectral weight at the 
Fermi level progressively increases upto $T_{IRM}$, corresponding to the Ioffe-Regel-Mott 
limit \cite{gunnarsson2003,hussey2004}.  The NFL regime can be inferred in terms of 
``resilient quasiparticles'', which continues to survive as the relevant low energy excitations in 
the system even after the Fermi liquid description breaks down \cite{deng2013}.

The breakdown of resilient QP description is signalled by two characteristics of 
$\sigma(\omega)$: $(i)$~ the DDP is broadened, undergoes suppression and gets monotonically shifted towards 
the high frequency range, $(ii)$~ the isosbestic crossing point of the $\sigma(\omega)$ curves is lost. 
We show these characteristics in Figure \ref{fig7}(f), where thermal evolution of $\sigma(\omega)$ is 
presented at a fixed disorder strength of $W=1.75t$. The isosbestic crossing point for this disorder is 
at $\omega \sim 5.8t$ and is lost for $T > 0.09t$ (see inset). The thermal evolution of $\sigma(\omega)$ 
thus encodes signature 
of resilient QP phase over the regime $T \le T_{IRM}$ and its subsequent crossover to the ``bad metallic'' 
phase for $T > T_{IRM}$. The crossover from the resilient QP to bad metal is a gradual degradation of 
the QP signatures and is not associated with any sharp transition. The QP energy scale given by the 
Brinkmann-Rice scale is tied to $T_{IRM}$ at which the QP description breaks down.  

\begin{figure}[b]
\begin{center}
\includegraphics[height=8.7cm,width=8.5cm,angle=0]{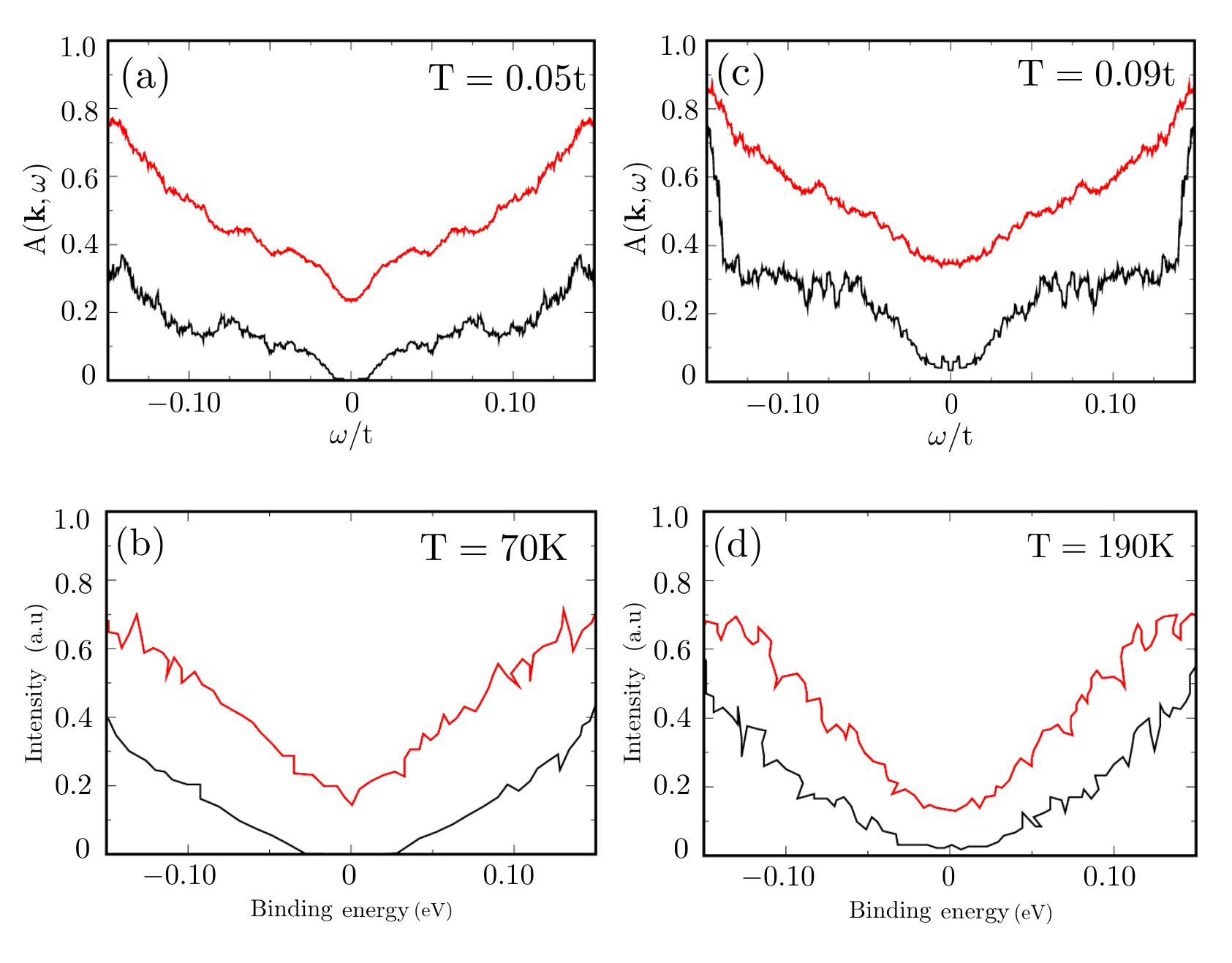}
\caption{Color online: Temperature dependence of numerically determined spectral lineshape 
((a) and (c)) and experimental EDC ((b) and (d)), for the clean (black)
and disordered (red) system. At low temperature the spectra is gapped out for the clean system while 
in presence of disorder there is a soft gap at the Fermi level. Temperature leads to monotonic 
closure of the gap via thermal fluctuations induced suppression of (quasi) long range angular 
correlation between the magnetic moments. Both the experimental and numerical data are symmetrized 
and for the numerical calculations the results correspond to $W = 1.75t$. The experimental data is 
extracted from Ref. \cite{trivedi2014}.}
\label{fig9}
\end{center}
\end{figure}

\section{Discussions}  
Resistivity measurements on Cu intercalated 1T-TaS$_{2}$ shows $d\rho/dT < 0$ in the 
metallic regime at low temperatures \cite{trivedi2014}, in agreement with our
results. Further experimental works on this and related class of systems are however called for, particularly
optical measurements which provide unambiguous evidence of NFL physics. For example, signatures of NFL state 
and resilient QPs observed in optical measurements has recently been reported in molecular charge transfer salt 
$\kappa$-[(BEDT-STF)$_{x}$(BEDT-TTF)$_{1-x}$]$_{2}$Cu$_{2}$(CN)$_{3}$, which undergoes chemical substitution 
induced IMT \cite{fratini2021}. Similarly, emergence of NFL metal via the interplay of quenched disorder and 
strong correlation has been recently observed in x-ray irradiated organic Mott insulator 
$\kappa$-(ET)$_{2}$Cu[N(CN)$_{2}$]Cl \cite{kanoda2020,kanoda2020_1}. 
The emergent novel electronic state close to the Mott boundary, as identified based on nuclear 
magnetic resonance (NMR) measurements, is suggested to be possibly the first experimental evidence of the 
much sought after ``electronic Griffiths phase'' \cite{andrade2009}.

In Cu intercalated 1T-TaS$_{2}$ disorder stabilized metallic regime has been reported 
based on the ARPES measurements \cite{trivedi2014}. In the Mott phase of clean 1T-TaS$_{2}$ the band lying 
closest to the Fermi energy has a bandwidth $W_{B} \approx 45 meV$, while the onsite Coulomb energy is 
found to be of the order of $\approx 0.1eV$ \cite{sipos2008}. Accordingly, $U/W_{B} \approx 2.2$, 
which is larger than $U_{c}/W_{B} \approx 1.3$ corresponding to the Mott transition and the system is 
well within the Mott insulating phase \cite{scalletar2006}. In the clean limit the single band Hubbard 
model on a triangular lattice is in the Mott insulating phase beyond $U_{c} \approx 5.27t$ \cite{rajarshi}. 
Our choice of $U = 6t$ in this work thus ensures that we are in the experimentally relevant regime for our 
numerical calculations.

\begin{figure}[t]
\begin{center}
\includegraphics[height=6.5cm,width=7.5cm,angle=0]{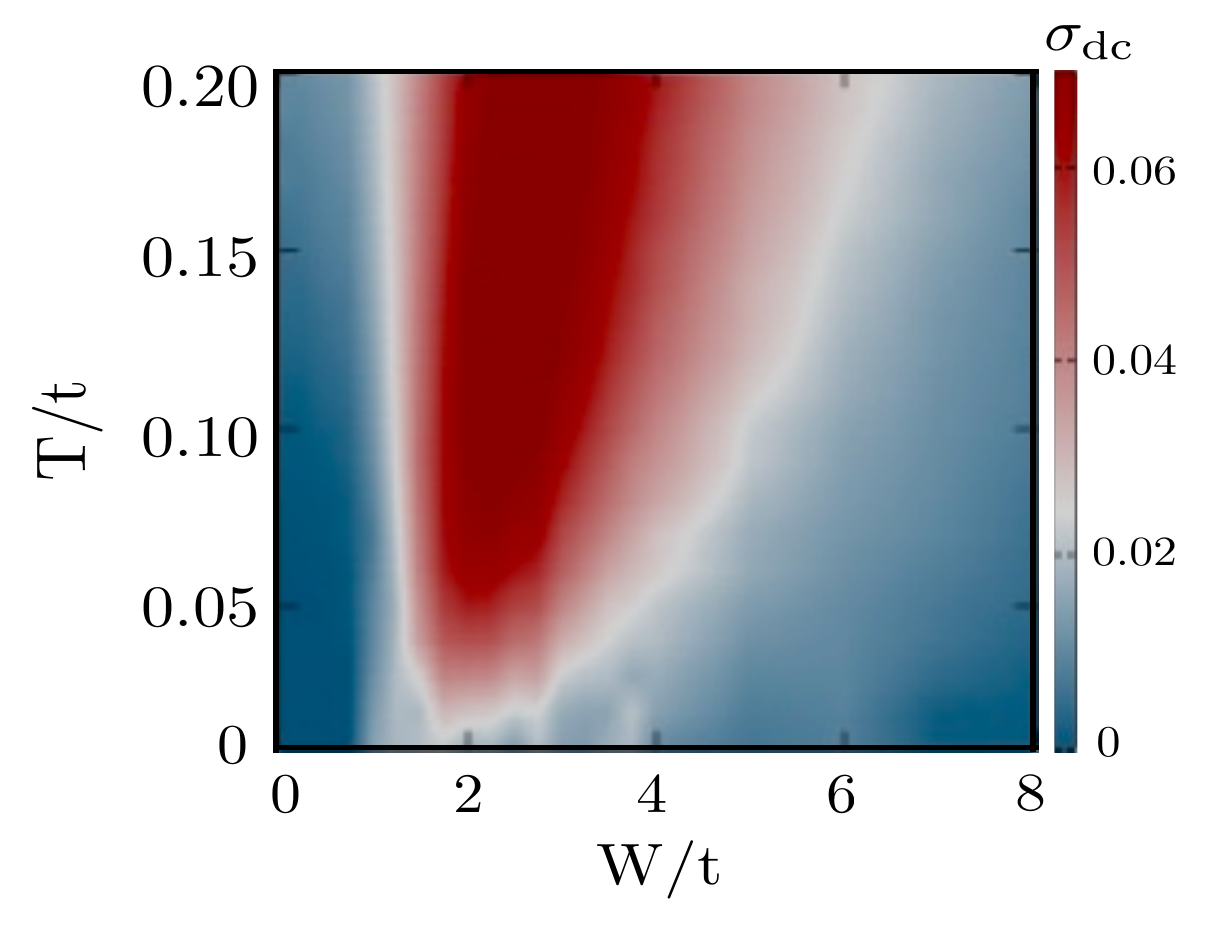}
\caption{Color online: Disorder stabilized NFL metal in the disorder-temperature ($W-T$)-plane 
depicted in terms of the $\sigma_{dc}$. The large conductivity at intermediate disorder regime 
(AFM) attests the metallic phase, while for the AF-MI and CAI regime at weak and strong disorder, respectively, 
the conductivity is strongly suppressed. Generic features of disorder tuned IMT shown in this figure 
can be compared with the experimental observations of applied electric field controlled interaction 
tuned IMT in 2D TMD frustrated Mott insulators \cite{chowdhury2021}.}
\label{fig10}
\end{center}
\end{figure}

The energy distribution curve (EDC) at the $\Gamma$-point of the 
Brillouin zone, as obtained from the ARPES measurements reveal that disorder leads to accumulation of 
spectral weight at the Fermi level and gives rise to pseudogap. The experimentally observed EDC at 
the $\Gamma$-point for clean and Cu intercalated 1T-TaS$_{2}$ are shown in Figure \ref{fig8}(a) \cite{trivedi2014}. 
For comparison we show the low temperature spectral lineshape $A({\bf k}, \omega)$ at the $\Gamma$-point, 
obtained numerically, in Figure \ref{fig8}(b). 
The disorder generated pseudogap at the Fermi level is in excellent 
agreement with the experimental observations. Thermal fluctuations lead to progressive closure of the pseudogap,  
accompanying the crossover of the system from the NFL to bad metallic phase, as is observed both experimentally 
as well as from our numerical calculations, is shown in Figure \ref{fig9}.

IMT and the associated NFL behavior discussed in this paper is a fairly generic feature 
of 2D TMDs. Recently signatures of NFL have been realized in TMD heterostructures such as,
MoTe$_{2}$/WSe$_{2}$ Moire superlattices \cite{chowdhury2021} and twisted bilayer WSe$_{2}$
\cite{pashupathy2021}. In Ref. \cite{chowdhury2021} continuous insulator-metal transition is realized in 
MoTe$_{2}$/WSe$_{2}$ Moire heterostructure at half filling via applied electric field tuned electronic 
interaction. The corresponding resistivity measurements show deviation from $T^{2}$ behavior 
thereby providing clear signature of NFL physics. The thermal phase diagram of this system as mapped 
out in the electric field-temperature ($E-T$)-plane in terms of the electrical resistance shows that a metallic 
phase is stablized close to the QCP (see Figure 3(c) of Ref. \cite{chowdhury2021}). The high temperature 
regime exhibits a crossover to the bad metallic phase. Moreover, magnetic 
susceptibility measurements confirmed the absence of long range magnetic order and/or presence of spiral 
antiferromagnetic order in this system at low temperatures. Notably, the loss of magnetic correlations is 
distinct from the IMT and gives rise to two different QCPs.

We establish the agreement of our results with the experimental observations reported 
in Ref. \cite{chowdhury2021} in terms of Figure \ref{fig10} (to be compared with Figure 3(c) of Ref. 
\cite{chowdhury2021}), wherein we map out the thermal phase diagram in the $W-T$ plane in terms of 
the dc conductivity $\sigma_{dc}$. The generic features captured by our calculations are in excellent 
agreement with those of the experimental observations on 2D TMD. This allows us to conclude that the numerical 
approach used in our work is versatile and is suitable to capture the generic features of IMT in 2D 
frustrated Mott insulators.

Note that the IMT and the emergent NFL metal are closely related to the spatial inhomogeneity
of the underlying phase. The role of spatial inhomogeneity was brought into focus by 
the advent of imaging techniques with nanoscale resolution, the most compelling 
example being VO$_{2}$ \cite{keilmann2007}. Near field infrared scanning spectroscopy on VO$_{2}$ 
thin films and microcrystals showed spatial proliferation of metallic and insulating ``puddles'' 
in presence of disorder, arising 
out of local fluctuations \cite{keilmann2007,basov2009,raschke2015,wolf2013}. 
The corresponding transport signatures are expectedly non trivial as the 
scattering properties now vary locally. 

Theoretical investigation of this rather complex phenomena of IMT requires one to take into account 
the fluctuations originating from disorder and thermal effects, a goal most of the powerful numerical techniques 
fail to achieve. In a recent work,  an extension of DMFT technique (statistical-DMFT) has been employed to address 
this issue on a square lattice, at finite temperatures \cite{miranda2020}. 
Geometric frustration further complicates the problem since standard numerical techniques such as DQMC cannot 
be used owing to the fermionic sign problem, while the Hartree-Fock mean field theory by construction is 
applicable at $T=0$ only. Our work presented in this paper provides an  
efficient numerical tool at reasonable computational expense. Access to large system sizes, real space correlators 
and real frequency dependent properties provide the edge to our approach over the existing ones.

In conclusion, we for the first time have provided the microscopic description of the disorder 
stabilized metal in 2D frustrated Mott insulators. Based on a non perturbative 
numerical technique we have mapped out the phases and the relevant thermal scales, exhibiting 
a disorder induced Mott insulator-metal transition. By analyzing the phases based on the thermodynamic, 
spectroscopic and transport signatures we have shown that the emergent metal is a non Fermi liquid,  
characterized by resilient quasiparticles and undergoes thermal crossover 
to bad metallic phase. As a test case for our approach we selected Cu intercalated 1T-TaS$_{2}$, which 
is a 2D transition metal dichalcogenide Mott insulator. Our results are in excellent agreement with the 
existing experimental observations on this material and are expected to serve as benchmarks for future 
experiments on this and related class of materials. Further, we demonstrated that the signatures of IMT 
is 2D frustrated Mott insulators are fairly generic and is well captured by our numerical approach. 

\section{Acknowledgements:-} 
The authors would like to thank the high performance computing cluster (HPCC) facility 
at Harish Chandra Research institute, Prayagraj (Allahabad), India. 

\section{Appendix}

\subsection{Numerical approach}

The Anderson-Hubbard model on a triangular lattice reads as \cite{fazekas_lecturenotes}, 
\begin{eqnarray}
H & = & \sum_{\langle ij\rangle, \sigma}t_{ij}(c_{i\sigma}^{\dagger}c_{j\sigma} + h . c)
+ \sum_{i,\sigma}(W_{i}-\mu)\hat n_{i\sigma} + U\sum_{i}\hat n_{i\uparrow}\hat n_{i\downarrow} \nonumber
\end{eqnarray}

\noindent here, $t_{ij}=-t$ for the nearest neighbor hopping on an isotropic triangular lattice. 
$t=1$ is set as the reference energy scale. $U > 0$ is the on-site repulsive Hubbard interaction. The site dependent
disorder is introduced in the system via $W_{i}$ which is randomly selected from an uniform distribution
$[+W/2, -W/2]$. We work at half filling and the chemical potential $\mu$ is adjusted to achieve the same.
In order to make the model numerically tractable we decompose the interaction term using Hubbard Stratonovich 
(HS) decomposition \cite{hubbard1959,schultz1990} and thereby introduce two (bosonic) auxiliary fields viz. a 
vector field ${\bf m}_{i}(\tau)$ and a scalar field $\phi_{i}(\tau)$, which couples to the spin and charge densities, 
respectively. The introduction of these auxiliary 
fields aid in to capture the Hartree-Fock theory at the saddle point, retains the spin rotation invariance 
and the Goldstone modes. In terms of the Grassmann fields $\psi_{i\sigma}(\tau)$, we have,
\begin{eqnarray}
\exp[U\sum_{i}\bar\psi_{i\uparrow}\psi_{i\uparrow}\bar\psi_{i\downarrow}\psi_{i\downarrow}] & = & \int {\bf \Pi}_{i}
\frac{d\phi_{i}d{\bf m}_{i}}{4\pi^{2}U}{\exp}[\frac{\phi_{i}^{2}}{U}+i\phi_{i}\rho_{i}+\frac{m_{i}^{2}}{U} 
\nonumber \\ && -2{\bf m}_{i}.{\bf s}_{i}]
\end{eqnarray}
where, the charge and spin densities are defined as, $\rho_{i} = \sum_{\sigma}\bar\psi_{i\sigma}\psi_{i\sigma}$ 
and ${\bf s}_{i}=(1/2)\sum_{a,b}\bar \psi_{ia}{\bf \sigma}_{ab}\psi_{ib}$, respectively. The corresponding 
partition function thus takes the form,
\begin{eqnarray}
{\cal Z} & = & \int {\bf \Pi}_{i}\frac{d\bar\psi_{i\sigma}d\psi_{i\sigma}d\phi_{i}d{\bf m}_{i}}{4\pi^{2}U}
\exp[-\int_{0}^{\beta}{\cal L}(\tau)]
\end{eqnarray}
where, the Lagrangian ${\cal L}$ is defined as,
\begin{eqnarray}
{\cal L}(\tau) & = & \sum_{i\sigma}\bar\psi_{i\sigma}(\tau)\partial_{\tau}\psi_{i\sigma}(\tau) + H_{0}(\tau) 
\nonumber \\ && +\sum_{i}[\frac{\phi_{i}(\tau)^{2}}{U}+(i\phi_{i}(\tau)+W_{i}-\mu)\rho_{i}(\tau)+
\frac{m_{i}(\tau)^{2}}{U} \nonumber \\ && -2{\bf m}_{i}(\tau).{\bf s}_{i}(\tau)]
\end{eqnarray}
where, $H_{0}(\tau)$ is the kinetic energy contribution. 
The $\psi$ integral is now quadratic but at the cost of an additional integration over
the fields ${\bf m}_{i}(\tau)$ and $\phi_{i}(\tau)$. The weight factor for the ${\bf m}_{i}$
and $\phi_{i}$ configurations can be determined by integrating out the $\psi$ and
$\bar \psi$;  and using these weighted configurations one goes back and computes
the fermionic properties. Formally,

{\begin{eqnarray}
{\cal Z} & = & \int {\cal D}{\bf m}{\cal D}{\phi}e^{-S_{eff}\{{\bf m},\phi\}}
\end{eqnarray}}
\begin{eqnarray}
S_{eff} & = & \log Det[{\cal G}^{-1}\{{\bf m},\phi\}] + \frac{\phi_{i}^{2}}{U} +
\frac{m_{i}^{2}}{U}
\end{eqnarray}
where, ${\cal G}$ is the electron Green's function in a $\{{\bf m}_{i},\phi_{i}\}$ background.

The weight factor for an arbitrary space-time configuration $\{{\bf m}_{i}(\tau), \phi_{i}(\tau)\}$ 
involves computation of the fermionic determinant in that background. The auxiliary field quantum Monte Carlo 
with static path approximation (SPA) retains the full spatial dependence
in ${\bf m}_{i}$ and $\phi_{i}$ but keeps only the $\Omega_{n}=0$ mode.
It thus includes classical fluctuations of arbitrary magnitudes but no quantum
($\Omega_{n} \neq 0$) fluctuations. 

Following the SPA approach, we freeze $\phi_{i}(\tau)$ to its saddle point value 
$\phi_{i}(\tau)=\langle n_{i} \rangle U/2 = U/2$ (at half filling). Note that this approximation is 
valid strictly at half filling where the charge fluctuations are suppressed, for
large $U$. Away from half filling, the charge fluctuations would be large even for large 
$U$. The resulting model can be thought of as fermions coupled to spatially fluctuating random 
background of classical field ${\bf m}_{i}$. With these approximations the effective Hamiltonian 
corresponds to a coupled spin-fermion model, which reads as, 
\begin{eqnarray}
H_{eff} & = & \sum_{\langle ij\rangle, \sigma}t_{ij}[c_{i\sigma}^{\dagger}c_{j\sigma}+h.c.] 
+ \sum_{i\sigma}(\frac{U}{2}+W_{i}-\mu)\hat n_{i\sigma} \nonumber \\ && - \frac{U}{2}\sum_{i}{\bf m}_{i}.{\bf \sigma}_{i} 
+ \frac{U}{4}\sum_{i}m_{i}^{2}
\end{eqnarray}
where, the last term corresponds to the stiffness cost associated with the now classical field 
${\bf m}_{i}$ and ${\bf \sigma}_{i}=\sum_{a,b}c_{ia}^{\dagger}{\bf \sigma}_{ab}c_{ib}={\bf s}_{i}$.
\begin{figure*}
\begin{center}
\includegraphics[height=10.4cm,width=14.5cm,angle=0]{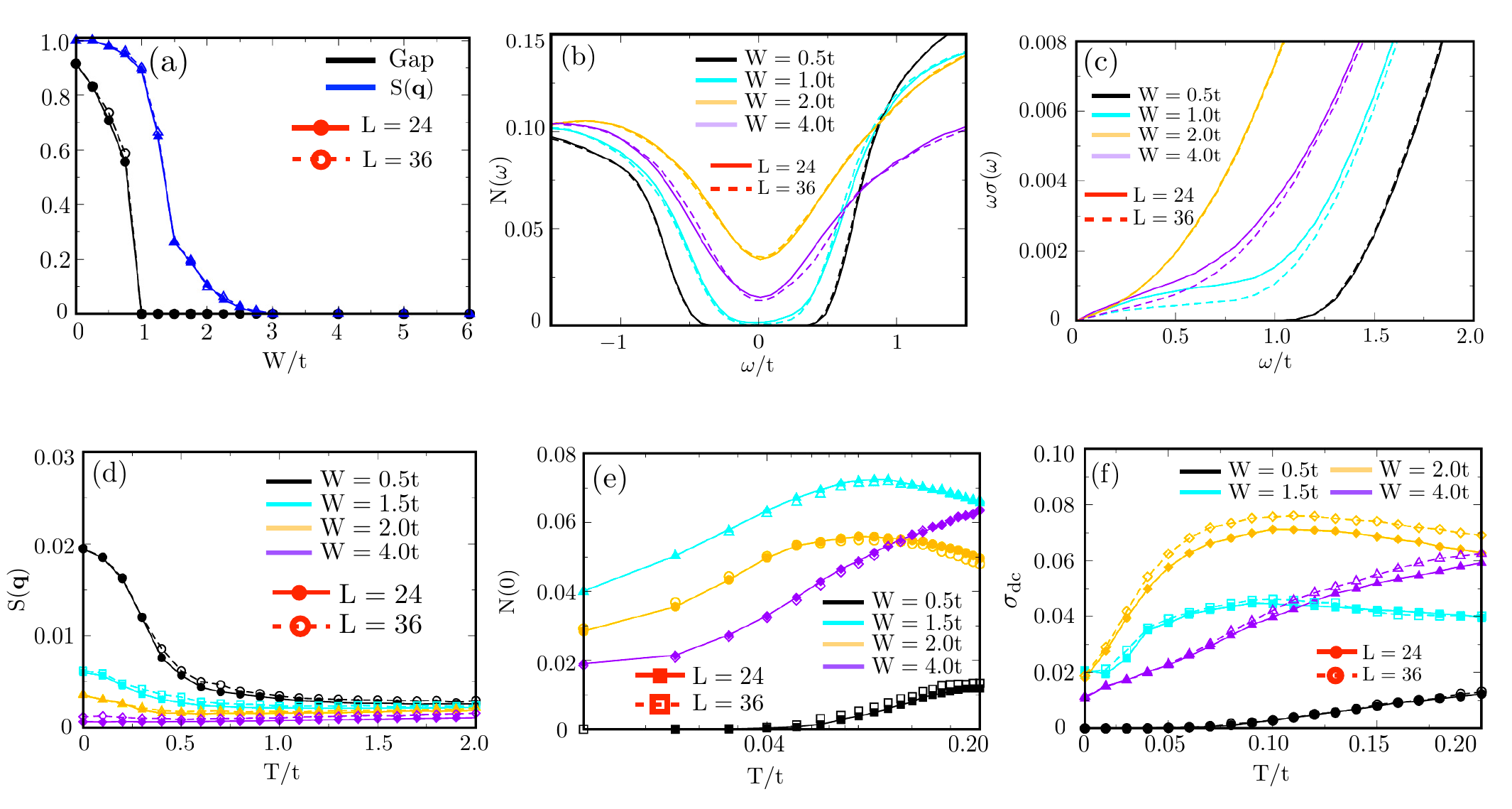}
\caption{Color online: Comparison of indicators at the ground state and at finite temperatures as calculated on 
system sizes of $L=36$ (dashed lines) and $L=24$ (solid lines). The panels correspond to,  (a) Mott gap and magnetic 
structure factor peak at $T=0$, (b) single particle DOS ($N(\omega)$) for different $W/t$ at $T=0$, (c) scaled optical 
conductivity ($\omega\sigma(\omega)$) in the low frequency regime for different $W/t$ at $T=0$, (d) temperature 
dependence of magnetic structure factor peak at ${\bf q} = \{2\pi/3,2\pi/3\}$ for selected $W/t$, (e) temperature 
dependence of single particle DOS ($N(0)$) at the Fermi level for different $W/t$, (f) temperature dependence of 
dc conductivity ($\sigma_{dc}$) at selected $W/t$.}
\label{fig11}
\end{center}
\end{figure*}

The random background configurations of $\{{\bf m}_{i}\}$ are generated numerically via Monte Carlo 
simulation and obey the Boltzmann distribution,
\begin{eqnarray}
P\{{\bf m}_{i}\} \propto Tr_{c,c^{\dagger}}e^{-\beta H_{eff}}
\end{eqnarray}
For large and random configurations the trace is computed numerically, wherein we diagonalize 
$H_{eff}$ for each attempted update of ${\bf m}_{i}$ and converge to the equilibrium configuration 
using Metropolis algorithm. Evidently, the process is numerically expensive and involves an 
${\cal O}(N^{3})$ computational cost per update (where $N=L\times L$ correspond to the system
 size), thus the cost per MC sweep is $N^{4}$. We cut down on the 
computation by using travelling cluster algorithm, wherein instead of diagonalizing the entire 
lattice for each attempted update of ${\bf m}_{i}$ we diagonalize a smaller cluster surrounding the 
update site \cite{sanjeev}. The computation cost now scales as ${\cal O}(NN_{c}^{3})$ (where $N_{c}$ 
is the size of a smaller cluster surrounding the update site), which is linear in lattice 
size $N$. This allows us to access large system sizes $\sim 40\times 40$ in two dimensions, which is 
essential to capture the inhomogeneity of the underlying magnetic phase. The equilibrium configurations 
obtained via the combination of Monte Carlo and Metropolis at different temperatures are used to compute 
the different fermionic correlators.  

SPA does not take into account the effect of quantum fluctuations. We do not 
expect a qualitative change in the results reported in this manuscript in case quantum fluctuations are
considered. For a clean system a {\it possible} consequence of taking into account the effect of 
quantum fluctuations is the restoration of the translation invariance of the lattice in the metallic phase. 
This is expected to make the metallic 
phase perfectly conducting at T=0 and above a low coherence temperature, signatures of a highly resistive metal 
would be observed. Inclusion of disorder in the system eliminates such a possibility because the translation 
symmetry of the lattice is anyway broken by disorder and cannot be restored by quantum fluctuations. In 
the presence of disorder spatial fluctuations play a vital role at finite temperatures.
\begin{figure*}
\begin{center}
\includegraphics[height=8.0cm,width=15cm,angle=0]{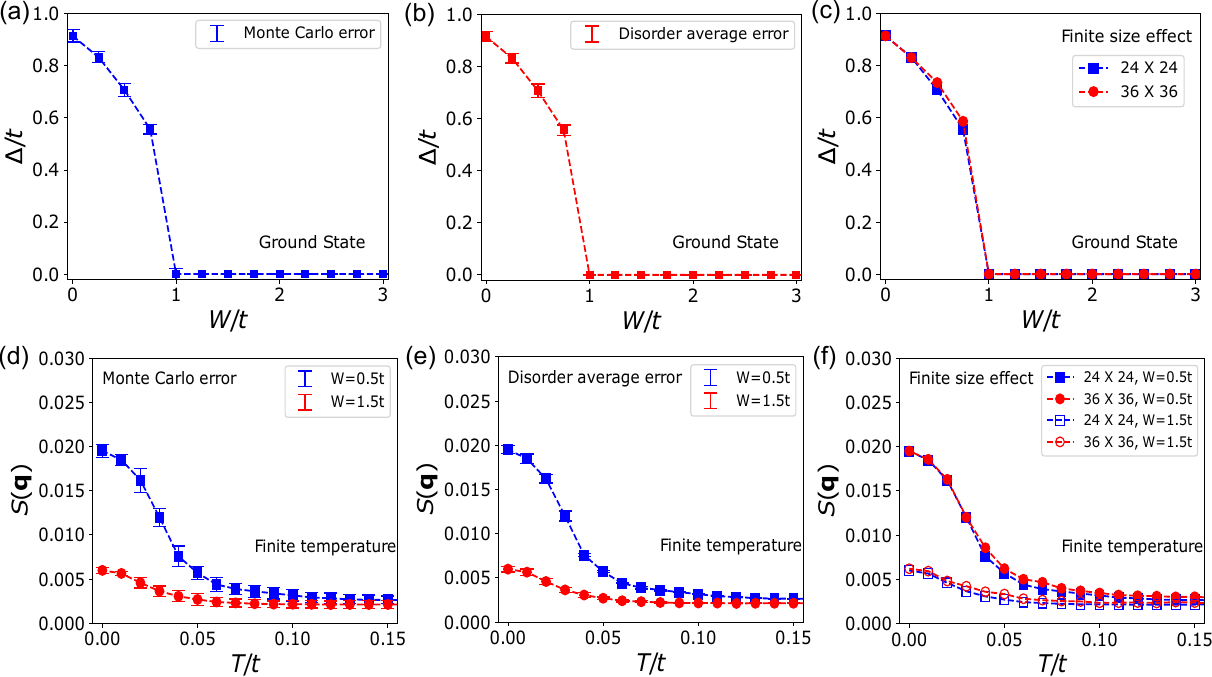}
\caption{Color online: Numerical error for representative indicators at the ground state and 
at finite temperatures, as selected disorder potentials. The error arising 
out of Monte Carlo sampling, disorder averaging and finite system size effects, 
are shown as separate panels, for the ground state and finite temperature indicators.}
\label{fig12}
\end{center}
\end{figure*}
 
Static path approximation has been used to investigate several quantum many body phenomena,
such as, BCS-BEC crossover in superconductors \cite{tarat_epjb}, 
Fulde-Ferrell-Larkin-Ovchinnikov (FFLO) superconductivity in solid 
state systems and ultracold atomic gases \cite{mpk_imb}, Mott transition in frustrated lattices 
\cite{rajarshi,nyayabanta_pyrochlore,nyayabanta_chk,nyayabanta_epl}, d-wave 
superconductivity \cite{dagotto_prl2005}, competition and coexistence of magnetic (AFM) and d-wave 
superconducting orders \cite{dagotto_prb2005}, orbital selective magnetism relevant for iron 
superconductors \cite{dagotto_prb2016}, strain induced superconductor-insulator 
transition in flat band lattices \cite{lieb_strain}, heteromolecular ultracold atomic gases \cite{mpk_mass} etc. 
In many of these problems the use of numerically exact techniques like DQMC is not feasible due either to the 
sign problem or to the system size restrictions (particularly for multiband systems). Judicial approximations 
are thus essential and SPA is one such approximation which can capture the ground state and thermal properties 
of these strongly correlated systems with reasonable accuracy. The technique however will and does fall short 
in situations where the physics of the ground state is almost entirely dictated by quantum fluctuations, such as, 
quantum spin liquids, heavy fermion superconductors etc.

\subsection{Finite system size effect}

The results discussed in the manuscript corresponds to a system size of $24 \times 24$. The choice of the 
system size is dictated by a balance between the computation cost and stability of the results obtained. 
Any lattice based simulation is however liable to be plagued by finite system size effects and in order to 
ascertain the robustness of our results against system sizes we have carried out the computation for selected 
disorder cross sections at a larger system size of $36 \times 36$. In Figure \ref{fig11} we show for the key 
indicators, the comparison between the results obtained for $L=24$ and $L=36$. We infer that our results are 
robust against the finite system size effects, both at the ground state and at finite temperatures.

In Figure \ref{fig12}, we systematically show the numerical error for the representative 
indicators at the ground state and at finite temperatures, arising out of, 
(a) Monte Carlo sampling, (b) disorder averaging and (c) finite system size effect. 
The data presented in our manuscript are averaged over 200 Monte Carlo configurations 
and 30 disorder realizations. In Figure \ref{fig12}, as the ground state indicator we show the 
disorder dependence of the Mott gap ($\Delta/t$), while as the finite temperature 
indicator the temperature dependence of the magnetic structure factor peak $S({\bf q})$ 
is presented at the selected disorder potential of $W=0.5t$ and $W=1.5t$. We have compared 
our results obtained at $L=24$ with those obtained at $L=36$, so as to ascertain the 
robustness of the thermodynamic phases and the phase boundaries shown in the manuscript. 
In principle, even larger system sizes can be accessed with our numerical approach but 
that involves larger computation cost.

Our results for the ground state and finite temperature indicators show that within 
our numerical scheme the largest contribution to numerical error comes from the Monte 
Carlo sampling, followed by the contribution from disorder averaging. For both contributions 
the errorbars are comparable to the size of the data points and it can be further reduced by 
averaging over a larger number of Monte Carlo samples and disorder realizations, respectively. 
Over the regime of weak and intermediate disorder potential our results are robust against 
finite system size effects. Figure \ref{fig12} shows that our choice of $L=24$ is sufficiently large 
to eliminate errors arising out of finite system sizes.

\subsection{First order phase transition and hysteresis}

\begin{figure}
\begin{center}
\includegraphics[height=6.0cm,width=6.3cm,angle=0]{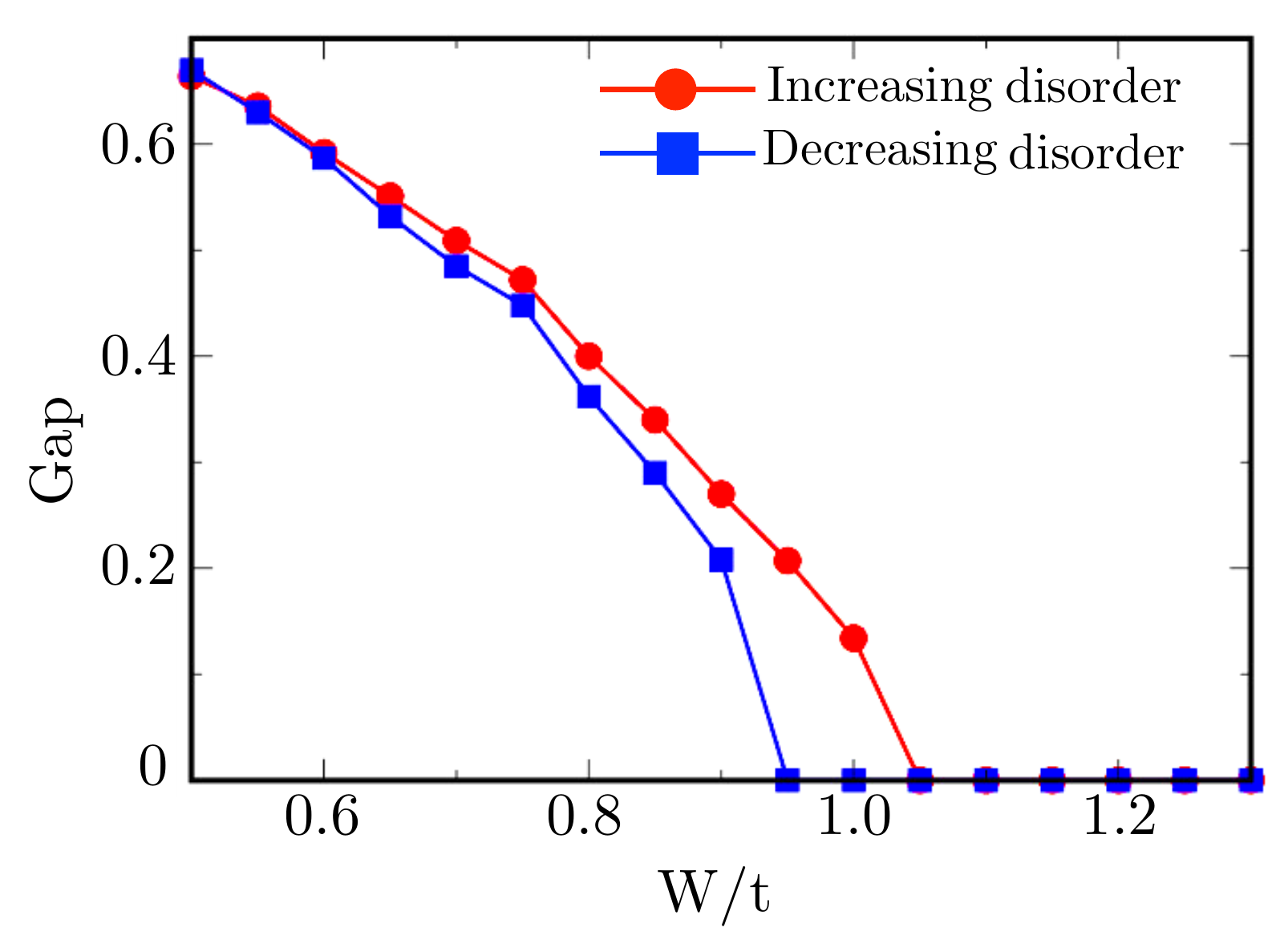}
\caption{Color online: Weakly first order phase transition of the Mott gap across the Mott insulator-metal 
transition. With increasing disorder the gap closes at $W = 1.05t$ while with decreasing disorder it closes 
at $W = 0.95t$.}
\label{fig13}
\end{center}
\end{figure}

We have shown in the main text that at the ground state the disorder tuned Mott insulator-metal 
transition is weakly first order in nature. The same has been depicted in term of the Mott gap as 
determined from the single particle DOS (see Figure \ref{fig2}(a) in the main text). An additional 
confirmation of the weak first order nature of this phase transition is obtained from the hysteresis 
behavior that the Mott gap shows as function of increasing and decreasing disorder, as presented in 
Figure \ref{fig13}. 

In order to verify the hysteresis of the spectral gap at the Fermi level we begin with the equilibrium 
configuration of the classical fields in the clean limit at the low temperature and subject it to a random 
disorder configuration with disorder potential strength $W$. A random number seed is used to generate this 
disorder configuration. The strength of the disorder potential is then progressively increased in steps of 
$\Delta W = 0.05t$ upto $W=3t$, while keeping the random number seed to be the same. A particular choice of 
random number seed corresponds to a single disorder realization. The process is repeated with different random 
number seeds (thus different disorder realizations) and for each realization the single particle DOS is calculated, 
which are then averaged over the disorder realizations. The averaged single particle DOS is used to determine the 
spectral gap at the Fermi level, as function of increasing disorder strength. 

Next, we consider the final equilibrium configuration of the classical fields at $W=3t$ for a single disorder 
realization as the input and progressively reduce the strength of the disorder potential in steps of $\Delta W = 0.05t$
till it reaches the clean limit. This process is repeated using the final equilibrium 
configurations of the classical fields at $W=3t$ for different disorder realizations. The disorder averaged single 
particle DOS for different disorder strength is then calculated and the corresponding spectral 
gap at the Fermi level is determined, as function of decreasing disorder strength. 
         
\bibliographystyle{apsrev4-1}
\bibliography{tr_mit.bib}

\end{document}